\documentclass[runningheads]{llncs}
\usepackage{graphicx}
\usepackage{subfigure}
\usepackage{amsmath,bm}
\usepackage{booktabs}
\usepackage{multirow}
\usepackage{multicol}
\usepackage{algorithm}
\usepackage{algorithmic}
\usepackage{multirow} 
\usepackage{array}
\usepackage{bm,bbm}
\usepackage{appendix}
\usepackage{flushend}
\usepackage{xspace}
\usepackage{amssymb}
\usepackage{hyperref}
\usepackage[misc]{ifsym}

\newsavebox\CBox

\begin{document}
\title{Cold-Start based Multi-Scenario Ranking Model for Click-Through Rate Prediction}
\titlerunning{Cold-Start based Multi-Scenario Ranking Model for CTR Prediction}
\author{
Peilin Chen\inst{1} \and
Hong Wen\inst{1} \and
Jing Zhang\inst{2} \and 
Fuyu Lv\inst{1} \and
Zhao Li\inst{3} \and
Qijie Shen\inst{1} \and
Wanjie Tao\inst{1}\and
Ying Zhou\inst{4} \and
Chao Zhang\inst{3}$^{(\textrm{\Letter})}$
}
%
\authorrunning{P. Chen et al.} 
%
\institute{
Alibaba Group, Hangzhou, China
\email{\{peilin.cpl,qinggan.wh,fuyu.lfy,qijie.sqj,wanjie.twj\}@alibaba-inc.com}\\ \and
The University of Sydney, Sydney, Australia\\
\email{chaimi.ustc@gmail.com}\\ \and
Zhejiang University, Hangzhou, China\\
\email{lzjoey@gmail.com,zczju@zju.edu.cn}\\ \and
Zhejiang Lab, Hangzhou, China\\
\email{zhouying@zhejianglab.com}
}

\maketitle              
\begin{abstract}

  Online travel platforms (OTPs), \emph{e.g.}, Ctrip.com or Fliggy.com, can effectively provide travel-related products or services to users. In this paper, we focus on the multi-scenario click-through rate (CTR) prediction, $i.e.$, training a unified model to serve all scenarios. Existing multi-scenario based CTR methods struggle in the context of OTP setting due to the ignorance of the cold-start users who have very limited data. To fill this gap, we propose a novel method named Cold-Start based Multi-scenario Network (CSMN). Specifically, it consists of two basic components including: 1) User Interest Projection Network (UIPN), which firstly purifies users’ behaviors by eliminating the scenario-irrelevant information in behaviors with respect to the visiting scenario, followed by obtaining users’ scenario-specific interests by summarizing the purified behaviors with respect to the target item via an attention mechanism; and 2) User Representation Memory Network (URMN), which benefits cold-start users from users with rich behaviors through a memory read and write mechanism. CSMN seamlessly integrates both components in an end-to-end learning framework. Extensive experiments on real-world offline dataset and online A/B test demonstrate the superiority of CSMN over state-of-the-art methods.

\end{abstract}
\section{Introduction}

With large-scale travel-related products and services available on the Online Travel Platforms (OTPs), $e.g.$, Ctrip.com or Fliggy.com, Click-Through Rate (CTR) prediction ~\cite{zhou2018deep,pi2020search,zhang2020empowering}, which aims at predicting the probability of users clicking items, has been playing an increasing role for delivering high-quality recommendation results and boosting the final platform revenues. Nowadays, the majority of CTR models are mainly built for single-scenario problems, $i.e.$, providing online service exactly for a scenario after trained only with data from the scenario. Here, a scenario refers to a specific spot where items are displayed, $e.g.$, \emph{Guess You Like} module in our app homepage. However, in many industrial applications of travelling recommendation, a user may be engaged with multiple travel scenarios, $e.g.$, \emph{Hot Spring}, \emph{Skiing}, $etc.$, where each scenario has its corresponding relevant candidate items to display, such as \emph{ski spots nearby} could be displayed to users who are visiting the \emph{Skiing} scenarios. In addition, the users in different scenarios also have different travel intentions such as \emph{leisure} for \emph{Hot Spring}, and \emph{adventure} for \emph{Skiing}. Consequently, multi-scenario CTR prediction, which is of great practical significance but has been largely under-explored, deserves more research efforts from academia and industry. 

A straightforward strategy is to build an individual model for each business scenario using the single-scenario CTR prediction methods, $i.e.$, DIN ~\cite{zhou2018deep}, DIEN ~\cite{zhou2019deep}, DIHN ~\cite{shen2022deep}. However, it has two apparent shortcomings: 1) training models for small-scale scenarios may suffer from severe data sparsity problem, and 2) maintaining different models for all business scenarios will be costly, which motivates us to devise a unified CTR prediction model to serve multiple scenarios simultaneously. Another possible strategy is to employ Multi-Task Learning (MTL) methods ~\cite{caruana1997multitask,ma2018modeling}, $i.e.$, one task for the corresponding scenario. However, we argue that MTL methods have significant difference from multi-scenario methods, where MTL methods simultaneously address various types of tasks in the same scenario, such as jointly predicting the CTR and conversion rate (CVR) tasks ~\cite{wen2019multi,wen2020entire,wen2021hierarchically}, while multi-scenario methods always focus on the same task, $i.e.$, CTR task, across multiple scenarios. What's more, it is very necessary for multi-scenario CTR methods to capture the scenario-shared information of various scenarios and explore scenario-specific characteristics simultaneously, where scenario-shared information means the overlapping users and candidate items among multiple scenarios, and scenario-specific characteristics indicates that users' interests with respect to different scenarios would be significantly different due to the data scale or topic-specific preferences among different scenarios. For facilitating following narration, we regard scenario-shared information and scenario-specific characteristics across various scenarios as the \emph{commonality} and \emph{discrimination} property, respectively. In fact, how to devise a unified and elaborate CTR prediction model for multiple scenarios is very challenging, especially exploiting the \emph{commonality} and \emph{discrimination} properties simultaneously.

To achieve this goal, several representative works towards the multi-scenario CTR prediction have been proposed. For example, STAR ~\cite{sheng2021one} trains a single model to serve all scenarios simultaneously by leveraging data from all scenarios, capturing users' interests effectively by employing shared centered parameters and scenario-specific parameters to exploit the \emph{commonality} and \emph{discrimination} property, respectively. SAR-Net ~\cite{shen2021sar} learns users' scenario-specific interests by harnessing the abundant data from different scenarios via two specific attention modules, leveraging the scenario features and item features to modulate users' behaviors for exploring the \emph{discrimination} property effectively. 
Meanwhile, it utilizes the bottom-shared embedding parameters to exploit the \emph{commonality} property. DADNN ~\cite{he2020dadnn} devises a unified model to serve multiple scenarios, where shared bottom block among all scenarios is employed to exploit the \emph{commonality} property, while scenario-specific head captures the characteristics of every scenario, $i.e.$, exploiting the \emph{discrimination} property. Although aforementioned methods have achieved remarkable performance for multi-scenario CTR prediction task with the consideration of the \emph{commonality} and \emph{discrimination} properties, they all neglect the cold-start issue which are high-frequently encountered in the OTPs setting. In practice, users' behaviors on OTPs are quite sparse or even absent compared with other e-commerce platforms since travel is a low-frequency demand, resulting in the cold-start issue and making it ineffective to learn the cold-start users' personalized preferences. How to tackle the cold-start issue in the context of multiple scenarios, especially considering the \emph{commonality} and \emph{discrimination} properties, has been unexplored and remains challenging.


To fully tackle the cold-start issue while exploiting the \emph{commonality} and \emph{discrimination} properties, in this paper, we propose a novel method named Cold-Start based Multi-scenario Network (CSMN). It consists of two fundamental components including a User Interest Projection Network (UIPN) and a User Representation Memory Network (URMN). Specifically, UIPN firstly purifies users’ behavior representations by eliminating the scenario-irrelevant information in behaviors with respect to the visiting scenario, then summarizes the purified behaviors with respect to the target item via an attention mechanism. In other words, even though a user has same and fixed behaviors, each behavior will obtain varied yet purified representation across different scenarios, followed by extracting user scenario-specific interest via attention mechanism, thus exploiting the \emph{discrimination} property. In addition, the resultant representation from UIPN component would be delivered to the URMN component, further being as a supplement cue to infer the interests of cold-start users. Specifically, URMN can make users with sparse behaviors benefit from users with rich behaviors through a memory read and write mechanism, where each slot in the memory can be regarded as a cluster of users who share similar profiles given target item at specific scenario, resulting in cold-start users can absorb well-purified interest representations from users with similar profiles yet rich behaviors, mitigating the cold-start issue effectively. Meanwhile, CSMN utilizes shared bottom block among all scenarios to address the \emph{commonality} property. The contributions of this paper is three-fold:

\begin{itemize}
    \item We propose a novel method named Cold-Start based Multi-scenario Network (CSMN) for multi-scenario CTR prediction task, 
    which facilitates learning users' distinct interest representations across multiple scenarios. 
    
    \item We devise two key components, including URMN and UIPN, to jointly address the aforementioned cold-start issue in the multi-scenario setting in an end-to-end manner, where the \emph{commonality} and \emph{discrimination} properties are effectively exploited for multi-scenario modelling.

    \item We conduct extensive experiments on both real-world offline dataset and online A/B test. The results demonstrate the superiority of the proposed CSMN over representative methods. CSMN now serves millions of users in our online travel platform, achieving 3.85\% CTR improvement.

\end{itemize}


\section{Related Work}


\textbf{CTR Prediction:} Recently, academia and industry communities have paid a lot of attention on CTR prediction not only from the perspective of feature interactions, $e.g.$, DCN ~\cite{wang2017deep}, NCF ~\cite{he2017neural}, DeepFM ~\cite{guo2017deepfm}, AutoInt ~\cite{song2019autoint}, but also from the perspective of users' sequential modelling, $e.g.$, DIN ~\cite{zhou2018deep}, DIEN ~\cite{zhou2019deep}, MIMN ~\cite{pi2020search}, DIHN ~\cite{shen2022deep}. Apart from these methods for single-scenario CTR prediction, multi-scenario CTR predictions also have drawn increasing attention, $i.e.$, training a unified model to serve all scenarios simultaneously. For example, STAR ~\cite{sheng2021one} employs shared centered parameters and scenario-specific parameters to learn users' interest representation across multiple scenarios. SAR-Net ~\cite{shen2021sar} learns users' scenario-specific interests by harnessing the abundant data from different scenarios via specific attention modules. However, both methods ignore the cold-start issue in the OTPs setting, thereby struggling in effectively discovering cold users’ real interests across multiple scenarios. By contrast, our proposed CSMN model leverages a User Representation Memory Network to interpret the interest of users with sparse behaviors from those of users with rich behaviors.

\textbf{Cold Start Recommendation:} Cold start issue has been widely recognized in representative recommender systems. Typically, there are three kinds of solutions to address it including: 1) resorting to more generalized auxiliary and contextual information ~\cite{barjasteh2016cold,li2019zero,chou2016addressing}; 2) cross-domain transfer (CDT) ~\cite{zhao2020catn,kang2019semi}, $i.e.$, users may have interactions with items in one domain while not in the other relevant domain. The goal of CDT is to effectively infer cold-start users’ preferences based on their interactions from one domain to the other relevant domain; and 3) meta-learning approaches ~\cite{lee2019melu,dong2020mamo}, which argue that users with similar inherent profiles should be recommended with similar items by leveraging users’ few behaviors. Despite effective, the SOTA methods for cold-start issue still struggle in the OTPs setting, since they do not address the cold-start issue in the context of multi-scenario CTR prediction. By contrast, our proposed CSMN can benefit users with sparse or even absent behaviors from users with rich behaviors through external memory mechanism, where each slot can be regarded as a cluster of users who share similar profiles, resulting in cold-start users can obtain interest representations from users with similar profiles yet rich behaviors.

\textbf{Multi-Task Learning:} Multi-Task Learning (MTL) ~\cite{ruder2017overview,ma2018modeling,tang2020progressive,caruana1997multitask} has been widely used in recommender systems, which benefits from the multi-objective optimization. For example, MMoE ~\cite{ma2018modeling} extends the efficient Mixture-of-Experts (MoE) shared-bottom structure to exploit a light-weight gating network to model the relationship of various tasks, which has been demonstrated to handle the task-specific information in a highly efficient manner. Going one step further, to address the seesaw phenomenon, PLE ~\cite{tang2020progressive} adopts a progressive routing mechanism to gradually extract and separate deeper semantic knowledge. In the context of the multi-scenario prediction task, it makes prediction for multiple scenarios towards the same task, $i.e.$, the CTR task, where the label spaces are same. Although we can build individual network for corresponding scenario on top of a shared-bottom structure, followed by employing classical MTL approaches for multi-objective optimization.
However, the consistency and discrepancy of various scenarios are coupled with each other tightly, resulting in the sophisticated relationships of multiple scenarios are difficult to disentangle. 
By contrast, we propose a User Interest Projection Network to disentangle scenario-specific interests from users' historical behaviors.

\section{The Proposed Approach}

\begin{figure*}
  \centering
  \includegraphics[width=0.95\textwidth]{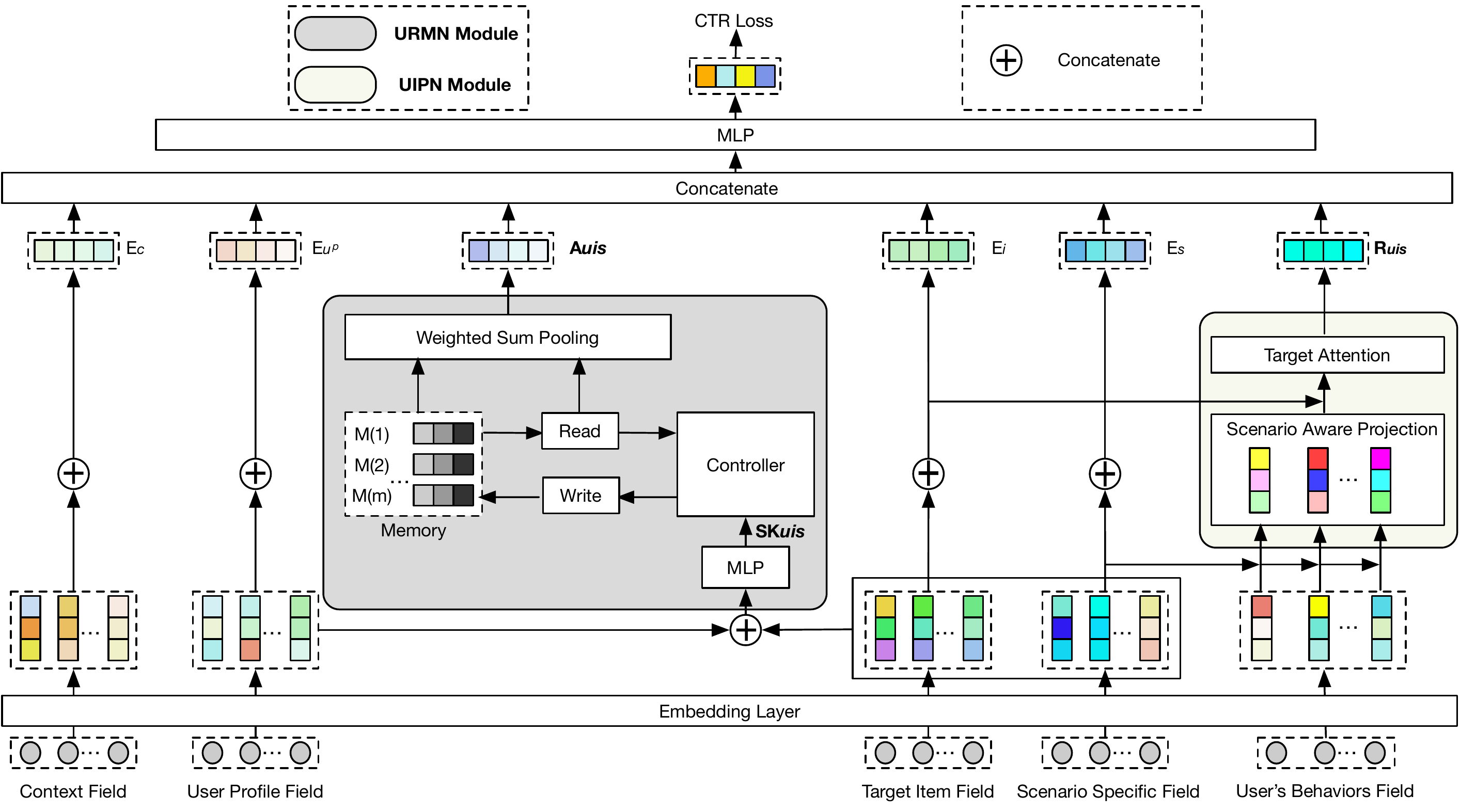}
  \caption{The overview architecture of the proposed CSMN model, which consists of Embedding Layer, User Representation Memory Network (URMN), and User Interest Projection Network (UIPN). Symbol $SK_{uis}$, $A_{uis}$, $R_{uis}$ denotes the comprehensive representation of the users’ profiles, target item and current visiting scenario, user’s augmented interest for the target item at specific scenario,  users' purified scenario-specific interests by leveraging the target items and current visiting scenarios, respectively.
  }
  \label{fig:model_arc}
\end{figure*}

In this paper, we propose a novel model named Cold-Start based Multi-scenario Network (CSMN) for multi-scenario CTR prediction. As depicted in Fig. \ref{fig:model_arc}, it consists of three basic components including Embedding Layer, User Interest Projection Network (UIPN), and User Representation Memory Network (URMN). We will introduce them in detail.

\subsection{Problem Definition}

In this section, we formally define the problem of multi-scenario CTR prediction task. Let $\mathcal{U}=\left \{ u_{1},u_{2},...,u_{N}  \right \}$, $\mathcal{I}=\left \{i_{1},i_{2},...,i_{M}  \right \}$, $\mathcal{S}=\left \{ s_{1},s_{2},...,s_{K}  \right \}$, $\mathcal{C}=\left \{ c_{1},c_{2},...,c_{L}  \right \}$ be a set of $N$ users, a set of $M$ items, a set of $K$ scenarios, a set of $L$ contexts, respectively. For facilitating the following narration, we omit the subscript and use symbols $u$, $i$, $s$, $c$ to denote user $u_{n}$, item $i_{m}$, scenario $s_{k}$, context $c_{l}$, respectively. And the user-item interaction at specific scenario is typically formulated as a matrix $Y=\left \{y_{uis}\right \}_{N\times M\times K}$. Specifically, $y_{uis}=1$ means user $u$ has clicked item $i$ at the scenario $s$, otherwise $y_{uis}=0$. In this paper, we mainly employ five types of input features namely \emph{User Profiles} $u^{P}$, \emph{User Behaviors} $u^{B}$, \emph{Target item} $i$, \emph{Context} $c$ and \emph{scenario} $s$ for each sample, where $u^{P}$ contains \emph{age}, \emph{sex}, \emph{purchase power}, $etc.$, $u^{B}$ denotes the sequential list of users visiting the set of items, $i$ contains \emph{item ID}, \emph{item's category ID}, $etc.$, $c$ contains \emph{weather}, \emph{time}, $etc.$, $s$ contains \emph{scenario ID}, \emph{scenario's accumulated CTR}, $etc.$, and $u$ can be defined as $\left \{ u^B;u^P \right \}$. Now, our learning goal is to train a unified CTR model to predict the probability $\hat{y}_{uis}$ of user $u$ clicks the target item $i$ at scenario $s$ given $u$, $i$, $s$, $c$, formulated as: $\hat{y}_{uis}=\mathcal{F} (u,i,s,c;\theta )$, where $\mathcal{F}$ and $\theta$ denote the learning objective and model parameters for the multi-scenario CTR prediction task, respectively.

\subsection{Embedding Layer}

Most data from industrial recommender systems are presented in a multi-field manner, where the fine-grained feature in each field is normally transformed into high-dimensional sparse one-hot features. For example, the one-hot vector representation of \emph{male} from the \emph{user sex} field can be decoded as $[1,0]^{T}$. Without loss of generality, we divide the raw data into five groups: target item $i$, user' historical behaviors $u^{B}$, user-specific profiles $u^{P}$, scenario information $s$ and context information $c$. Assuming the concatenation results of different fields’ one-hot vectors from these five groups as $X_{i}$, $X_{u^{B}}$, $X_{u^{P}}$, $X_{s}$ and $X_{c}$, respectively, they can be further transformed into low dimensional dense representations by multiplying corresponding embedding matrices, denoted as $E_{i}$, $E_{u^{B}}$, $E_{u^{P}}$, $E_{s}$ and $E_{c}$, respectively, $e.g.$, $E_{u^{B}}=[e_{1};e_{2};...;e_{T}]$, where $T$ and $e_{t}$ represent the length of users’ behaviors and the embedding feature of the $t$-th behavior, respectively. In this paper, we also employ bottom-shared embedding parameters to exploit the \emph{commonality} property. Since the bottom-shared embedding parameters make up the majority of the trainable parameters, they can be learned sufficiently by information sharing among overlapping users and candidate items, thereby avoiding the overfitting issue.
 
\subsection{User Interest Projection Network}

Generally, users' interests can be effectively extracted from users' historical behaviors. For example, as an excellent representative method for users' interest extraction, DIN ~\cite{zhou2018deep} firstly employs the attention mechanism to dynamically compute the weight of users' historical behaviors with respect to different target items, followed by utilizing a weighted-sum pooling operation to adaptively generate users' interest representation. Despite effective, we figure out it is not directly suitable for the multi-scenario CTR task in the OTPs setting, where not only the target item but also the scenario-specific information can affect users' interests. For example, when a user comes into two scenarios with varied topics, $e.g.$, \emph{Hot Springs} and \emph{Skiing} scenarios, user's interest can be significantly different since the user may be concerned about the \emph{leisure} in the \emph{Hot Springs} scenario while preferring \emph{outdoor adventure} in the \emph{Skiing} scenario.

Specifically, we argue that even though given the same target item and fixed user's behaviors, the representations of users' behaviors with respect to different scenarios will be inevitably varied. Therefore, a straightforward strategy towards the multi-scenario CTR prediction is to disentangle the scenario-specific characteristics from users' behaviors with respect to the current visiting scenario, which motivates us to propose the User Interest Projection Network (UIPN) module. Specifically, the representation of each element in $E_{u^{B}}$ will be projected into the orthogonal space of the scenario embedding $E_{s}$ to eliminate the scenario-irrelevant information. Formally, without loss of generality, we illustrate the orthogonal mapping process with a randomly selected element $e_{i}$ from $E_{u^{B}}$ and the dense feature $E_{s}$ of scenario $s$. Firstly, we embed $e_{i}$, $E_{s}$ into the same space by multiplying individual mapping matrix $W_{o}$, $W_{s}$, respectively, $i.e.$, $f_{i}=W_{o}e_{i}$, $f_{s}=W_{s}E_{s}$. Then, the refined preference representation vector $f_{i}^{p}$ with respect to scenario $s$ can be obtained by projecting the vector $f_{i}$ onto the direction of vector $f_{s}$, defined as $f_{i}^{p}=project(f_{i}, f_{s})$, where $project(,)$ denotes the scenario aware projection operator, $i.e.$, $project(a,b)=\frac{ab}{|b|}\frac{b}{|b|}$. $|\cdot|$ denotes the norm of a vector. 
In this manner, the original $E_{u^{B}}$ can be formulated as $f_{u^{B}}=\left \{ f_{1}^{p};f_{2}^{p};...;f_{T}^{p}  \right \}$. Inspired by the Multi-Head Self-Attention (MHSA) mechanism ~\cite{vaswani2017attention}, which can effectively capture the dependency between any pair within the sequence despite their distance, we use it to further enhance the representation of users' preference. Specifically, given $f_{u^{B}}=\left \{ f_{1}^{p};f_{2}^{p};...;f_{T}^{p}  \right \}$, we can obtain the enhanced representation $f_{u^{B}}^{'}=\left \{ f_{1}^{'p};f_{2}^{'p};...;f_{T}^{'p} \right \}$ after applying MHSA on $f_{u^{B}}$.

Next, to obtain users' purified interests from $f_{u^{B}}^{'}$, we need to calculate the similarity between the target item $i$ and each element of $f_{u^{B}}^{'}$, which can be formulated as $\alpha _{t}=Relu(z^{T}tanh(W_{i}E_{i}+W_{f}f_{t}^{'p}+b))$, 
where, $z$, $W_{i}$, $W_{f}$, and $b$ are all learnable parameters. After normalization, $i.e.$, $\alpha _{t}=\frac{exp(\alpha _{t})}{\sum_{i=1}^{T}exp(\alpha _{i}) } $, denoting the weight for the $t$-th behavior with respect to the target item $i$. Therefore, the final user' purified interest representation \textbf{$R_{u}$} from UIPN can be calculated as $R_{uis}=\sum_{i=1}^{T}\alpha_{i}f_{i}^{'p} $ via the weighted-sum pooling operation. In this way, UIPN can effectively achieve the interests of users who have rich historical behaviors with respect to current visiting scenario and the target item within it, which can be further exploited as a supplement cue to infer the interests 
of cold-start users who have similar profiles with rich behavior users.


\subsection{User Representation Memory Network}

In the OTP settings, users' behaviors are quite sparse compared with other typical e-commerce platforms, $i.e.$, resulting in the cold-start issue and making it difficult to extract users' interests from their behaviors. However, we argue that users' interests not only can be reflected from their behaviors but also from their inherent profiles. In other word, users' behaviors can be regarded as the embodiment of their inherent profiles. For example, when providing online service for a user with \emph{adventure spirit}, we can probably infer the user prefers \emph{Skiing} more than \emph{Hot Springs}, even though the user has no any online behaviors before. Therefore, when users' historical behaviors are sparse or even absent, users' inherent profiles can act as a kind of supplementary cues to discover users' interests. However, how to effectively extract users' interests from their profiles within the unified framework is nontrivial. To this end, we propose a User Representation Memory Network (URMN).

Specifically, URMN customizes an augmented vector $A_{uis}$ for each sample to represent user's augmented interest for the target item at specific scenario, followed by concatenating it with other representation features, together for model training. To obtain $A_{uis}$, we borrow the idea from Neural Turing Machine ~\cite{graves2014neural} which can store information in a fixed size of external memory. Specifically, we firstly generate a specific key $SK_{uis}$, which can be regarded as the comprehensive representation of the users' profiles, target item, and current visiting scenario. Then, we traverse all the slots of the external memory in URMN and generate each slot's weight with respect to the specific key $SK_{uis}$. Finally, we achieve the augmented vector $A_{uis}$ by a weighted-sum memory summarization. We will detail them as follows. 

First, the specific key $SK_{uis}$ defined as $SK_{uis}=F(\sum_{j=1}^{P}w_{j}e_{p_{j}};E_{i};E_{s})$,
where $P$ denotes the number of user' profiles, $F(.)$ represents three MLP layers with ReLU activation function, $w_{j}$ is the weight of user's profile $e_{p_{j}}$ with respect to the target item $E_{i}$ via attention mechanism. Intuitively, similar keys $SK_{uis}$ cluster together, implying that given specific scenario and target item, users having similar profiles probably share the similar interests, thus cold start users can benefit from users with similar profiles yet rich behaviors. From another perspective, all the learning parameters in $F(.)$ are shared, $e.g.$, shared MLP parameters, which also implies the representation of current key can be affected by the representations of other keys.

Next, we detail the structure of the memory in URMN with its parameters denoted as $Mem$. It consists of $q$ memory slots $\left \{ Mem_{i} \right \}|_{i=1}^{q}  $ with each slot containing corresponding key $Mem_{i}^{Key}$ and value $Mem_{i}^{Value}$, $i.e.$, $Mem_{i}\triangleq\left \{  Mem_{i}^{Key} , Mem_{i}^{Value} \right \}$. Each slot can be regarded as a cluster, where $Mem_{i}^{Key}$ (resp. $Mem_{i}^{Value}$) is updated by itself and $SK_{uis}$ (resp. $R_{uis}$). Moreover, $R_{uis}$ from the resultant representation of UIPN component denotes users' purified interests, depicted in Fig. \ref{fig:model_arc}. Specifically, two basic operations of URMN are \emph{Memory Read} and \emph{Memory Write}, which interact with memory through a controller.

\textbf{Memory Read}: During the \emph{Memory Read} process, the controller generates a read key $SK_{uis}$ as mentioned above to access the memory. Formally, it can be formulated as follows: $w_{uis}^{j}=\frac{exp(F_{xy}(SK_{uis},Mem_{j}^{Key}))}{\sum_{j=1}^{q}exp(F_{xy}(SK_{uis}, Mem_{j}^{Key})) } , j=1,...,q$,
where $F_{xy}(x,y)=\frac{x^{T}y }{\left \| x \right \| \left \| y \right \|}$, $w_{uis}^{j}$ is the weight of $SK_{uis}$ with respect to the key of slot $j$. Then, we obtain user's augmented interest vector $A_{uis}$ by weighted-sum pooling, defined as: $A_{uis}=\sum_{j=1}^{q}w_{uis}^{j}Mem_{j}^{Value}$.


\textbf{Memory Write}: Using the same weight $w_{uis}^{j}$, the update process of memory key and value is defined as follows:
\begin{equation}
Mem_{j}^{Key}=\alpha _{k}w_{uis}^{j}SK_{uis} + (1-\alpha _{k})Mem_{j}^{Key} ,
\label{eq:memory_write_key}
\end{equation}
\begin{equation}
Mem_{j}^{Value}=\alpha _{v}w_{uis}^{j}R_{uis} + (1-\alpha _{v})Mem_{j}^{Value} ,
\label{eq:memory_write_key}
\end{equation}
where $\alpha _{k}$ and $\alpha _{v}$ are the hyper-parameters valued in $[0,1]$, controlling the update rate of each memory slot's key and value, respectively. $Mem_{j}^{Key}$ and $Mem_{j}^{Value}$ are randomly initialized. In this way, URMN can distribute the interest of users with rich behaviors to the users with sparse or absent behaviors in the same cluster, alleviating the cold-start issue effectively.

Finally, all the representation vectors including $E_{i}$, $E_{u^{P}}$, $E_{s}$, $E_{c}$, $A_{uis}$ and $R_{uis}$ are concatenated, followed by feeding them to multiple MLP layers to generate the final predicted probability $\hat{y}_{uis}$. Now, given the predicted $\hat{y}_{uis}$ and the ground truth $y_{uis}\in \left \{ 0,1 \right \}$, we define the objective function as the negative log-likelihood function, formulated as: 

\begin{equation}
Loss=-\frac{1}{Num}\sum (y_{uis}log\hat{y}_{uis} + (1-y_{uis})log(1-\hat{y}_{uis})) ,
\label{eq:loss}
\end{equation}

where, $Loss$ is the total loss and $Num$ denotes the number of training data collected from all the scenarios.



\section{Experiments}

In this section, we conduct extensive offline and online experiments to comprehensively evaluate the effectiveness of the proposed CSMN, and try to answer following questions: 
\begin{itemize}
    \item \textbf{Q1}: How about the overall performance of the proposed CSMN compared with state-of-art methods? 
    \item \textbf{Q2}: How about the impact of each component of the proposed CSMN?
    \item \textbf{Q3}: How about the influence of key hyper-parameters, $e.g.$, the number of URMN slots? 
    \item \textbf{Q4}: How about the online performance of the proposed CSMN compared with other methods?
\end{itemize}

\subsection{Experiments Settings}
\subsubsection{Dataset Description}

To the extent of our knowledge, there are no public datasets suited for the multi-scenario CTR prediction task in the OTP settings. We make the offline dataset by collecting users' traffic logs from our OTP platform, which contain 29 million users and 0.61 million travel items from 20 scenarios in consecutive 30 days, $i.e.$, from 2022-05-22 to 2022-06-20. They are further divided into the disjoint training set and testing set, where the training set is from 2022-05-22 to 2022-06-19, while the testing set is from the left days. The statistics of this offline dataset are listed in Table \ref{pic:dataset_des}, where \emph{CSU Ratio} representing cold-start users' ratio. We can find that the data scales and distributions among these travel scenarios are significantly different. In addition, we observe that over 28\% (resp. 40\%) of users do not have any behaviors in recent 180 (resp. 90) days from the training data, which indeed implies the cold-start issue.

\begin{table}[]
  \caption{The statistics of the offline dataset.}
  \label{pic:dataset_des}
  \scalebox{0.98}{
  \begin{tabular}{cllllllllll}
    \toprule
   Scenario        & \#1     & \#2    & \#3    & \#4    & \#5    & \#6    & \#7    & \#8    & \#9    & \#10 \\
    \midrule
    \#User & 2.3M & 2.1M & 8.2M & 0.6M & 3.4M & 0.4M & 0.3M & 0.3M & 2.5M & 0.2M\\
    \#Item & 53K & 37K & 122K & 23K & 37K & 51K & 33K & 18K & 86K & 63K\\
    CTR        & 2.57\%  & 1.22\% & 1.64\% & 5.54\% & 8.10\% & 1.27\% & 6.75\% & 2.26\% & 11.61\% & 6.51\% \\
    CSU Ratio        & 27.65\%  & 23.92\% & 11.49\% & 18.66\% & 16.49\% & 31.28\% & 14.54\% & 22.93\% & 28.73\% & 38.52\%\\
    \hline
    \hline
   Scenario & \#11   & \#12   & \#13   & \#14   & \#15   & \#16   & \#17   & \#18    & \#19   & \#20   \\
    \hline
   \#User  & 1.1M & 0.6M & 0.6M & 1.7M & 9.1M & 0.17M & 0.16M & 0.14M & 0.13M & 0.11M \\
   \#Item & 135K & 41K & 57K & 27K & 11K & 5K & 66K & 23K & 31K & 16K \\
   
  CTR & 18.70\% & 9.87\% & 4.05\% & 14.38\% & 5.80\% & 1.72\% & 8.02\% & 4.23\%  & 3.03\% & 1.62\%\\
  
  CSU Ratio & 32.69\% & 16.77\% & 28.46\% & 19.55\% & 27.44\% & 26.22\% & 25.37\% & 10.45\%  & 30.07\% & 34.76\%\\

  \bottomrule
\end{tabular}}
\label{pic:dataset_des}
\end{table}

\subsubsection{Competitors}

To verify the effectiveness of the proposed CSMN, we compare it with following methods:
\begin{itemize}
\item \textbf{WDL} ~\cite{cheng2016wide}: It consists of wide linear and deep neural parts, which combines the benefits of memorization and generalization for CTR prediction.

\item \textbf{DeepFM} ~\cite{guo2017deepfm}: It imposes a factorization machine as a ``wide" part in WDL to eliminate feature engineering.

\item \textbf{DIN} ~\cite{zhou2018deep}: It extracts users' dynamic interest from their historical behavior via attention mechanism.

\item \textbf{MMOE} ~\cite{ma2018modeling}: It models the relationship among different tasks by employing gating networks and multi-task learning framework. We also adapt MMoE for multi-scenario task by assigning each output for corresponding scenario.

\item \textbf{PLE} ~\cite{tang2020progressive}: It contains shared components and task-specific components, and adopts a progressive routing mechanism to extract and separate deeper knowledge, enabling the efficiency of representation across multiple tasks. 

\item \textbf{STAR} ~\cite{sheng2021one}: It trains a unified model to serve all scenarios simultaneously, containing shared centered parameters and scenario-specific parameters. 

\item \textbf{SAR-Net} ~\cite{shen2021sar}: It predicts users’ scenario-specific interests from scenario/target item features and adaptively extracts scenario-specific information across multiple scenarios.

\end{itemize}

\subsubsection{Metrics and Implementation Details} To comprehensively evaluate the performance of different methods, we adopt two widely used metrics in recommender systems, $i.e.$, Area Under Curve (AUC) ~\cite{zhou2018deep,shen2022hierarchically} and Relative Improvement (RI) ~\cite{shen2021sar}, where, the larger AUC means better ranking performance, and RI provides an intuitive comparison measure by calculating the relative improvement of a target model over the baseline model. In addition, the proposed CSMN and other competitors are implemented by distributed Tensorflow 1.4, where learning rate, mini-batch, and optimizer, are set as 0.001, 1024, Adam ~\cite{kingma2014adam}, respectively. In addition, there are 4 layers in the MLP. Logistic loss is used as the loss function for all the competitors, as summarized in Table~\ref{tab:parameter}.

\begin{table}[t]
\small
    \caption{Hyper-parameters of all competitors. }
    \setlength\tabcolsep{17.0pt}
    \centering
    \label{tab:parameter}
    \begin{tabular}{c|cc}
    \toprule
    Hyper-parameters & Choice\\
    \midrule
    Loss function& Logistic Loss\\
    Optimizer & Adam \\
    Number of layers in MLP & 4\\
    Dimensions of layers in MLP & [512,256,128,32] \\
    Batch size & 1024\\
    Learning rate& 0.001 \\
    Dropout ratio& 0.5 \\
  \bottomrule
    \end{tabular}
\end{table}

\subsection{Experimental Results (Q1)}

In this subsection, we report the AUC results of all the competitors on the offline test set. As illustrated in Table \ref{tab:overallresult}, the consistent improvement of the proposed CSMN over other competitors validates its effectiveness. It achieves the best AUC results in each single scenario. Note that compared with WDL, DeepFM, DIN, MMOE and PLE, the multi-scenario CTR methods, $e.g.$, STAR, SAR-Net and the proposed CSMN, consistently achieves better performance, demonstrating that ignoring the scenario difference during the extraction of users' scenario-specific interests will seriously degenerate the performance of multi-scenario CTR prediction models. Nevertheless,  SAR-Net  still struggles in extracting users’ real interests across multiple different scenarios, since it cannot address the cold-start issue in the OTPs setting. By contrast, our CSMN leverages the URMN to specifically mitigate the adverse effect of them, respectively. Consequently, it achieves an improvement of 2.41\% RI over SAR-Net. Moreover, STAR also neglects the cold-start problem in the OTPs setting. Therefore, it has worse performance than our CSMN, especially for users with sparse (or even absent) behaviors. For example, for scenario \#10 and scenario \#20, which have very large portion of cold-start users, $i.e.$, over 38.52\% and 34.76\% respectively, depicted in Table \ref{pic:dataset_des}, CSMN achieves a larger AUC improvement of 3.11\% and 2.93\% over SAR-Net ,respectively. These results demonstrate the effectiveness of the proposed URMN in dealing with the cold-start issue, which clusters users according to the representation of their profiles and obtains the interest of users with sparse behaviors from neighboring users with rich behaviors. 
\begin{table}
\small
  \caption{The results of all methods on the offline dataset.}
  \label{tab:overallresult}
  \centering
   \resizebox{.85\columnwidth}{!}{
  \begin{tabular}{c|c|c|c|c|c|c|c|c|c}
    \toprule
scenario  & WDL &DeepFM & DIN &MMOE & PLE & STAR & SAR-Net & \textbf{CSMN} & RI\\
\midrule
\texttt{\#1}&0.6317 &0.6342 & 0.6368 & 0.6504 & 0.6506 & 0.6513 & \underline{0.6527} & \textbf{0.6546} & 1.24\%  \\
\texttt{\#2}&0.6695&0.6710 & 0.6718 & 0.6897 & 0.6901 & 0.6925 & \underline{0.6933} & \textbf{0.6959} & 1.35\%  \\
\texttt{\#3}&0.7186&0.7228 & 0.7231 & 0.7306 & 0.7307 & \underline{0.7328} & 0.7321 & \textbf{0.7331} & 0.13\% \\
\texttt{\#4}&0.6500&0.6511 & 0.6559 & 0.6678 & 0.6681 & 0.6709 & \underline{0.6714} & \textbf{0.6738} & 1.40\%\\
\texttt{\#5}&0.6822&0.6861 & 0.6882 & 0.7018 & 0.7016 & 0.7057 & \underline{0.7063} & \textbf{0.7079} & 0.78\% \\
\texttt{\#6}&0.6257&0.6304 & 0.6322 & 0.6482 & 0.6485 & \underline{0.6501} & 0.6496 & \textbf{0.6541} & 2.66\% \\
\texttt{\#7}&0.6388&0.6419 & 0.6447 & 0.6501 & 0.6522 & 0.6593 & \underline{0.6602} & \textbf{0.6609} & 0.44\% \\
\texttt{\#8}&0.6535&0.6533 & 0.6575 & 0.6624 & 0.6626 & \underline{0.6631} & 0.6628 & \textbf{0.6654} & 1.41\%\\
\texttt{\#9}&0.7032&0.7037 & 0.7058 & 0.7126 & 0.7131 & 0.7138 & \underline{0.7165} & \textbf{0.7186} & 0.97\% \\
\texttt{\#10}&0.6867&0.6921 & 0.6913 & 0.7003 & 0.7001 & 0.7012 & \underline{0.7024} & \textbf{0.7087} & 3.11\% \\
\texttt{\#11}&0.6701&0.6757 & 0.6772 & 0.6795 & 0.6802 & 0.6815 & \underline{0.6822} & \textbf{0.6864} & 2.31\% \\
\texttt{\#12}&0.7025&0.7042 & 0.7016 & 0.7163 & 0.7169 & 0.7176 & \underline{0.7189} & \textbf{0.7205} & 0.73\% \\
\texttt{\#13}&0.7258&0.7291 & 0.7294 & 0.7431 & 0.7438 & 0.7460 & \underline{0.7468} & \textbf{0.7504} & 1.46\% \\
\texttt{\#14}&0.6662&0.6708 & 0.6720 & 0.6789 & 0.6800 & 0.6813 & \underline{0.6862} & \textbf{0.6884} & 1.18\% \\
\texttt{\#15}&0.7135&0.7138 & 0.7131 & 0.7319 & 0.7323 & 0.7342 & \underline{0.7347} & \textbf{0.7399} & 2.22\% \\
\texttt{\#16}&0.6411&0.6532 & 0.6560 & 0.6698 & 0.7005 & 0.7019 & \underline{0.7025} & \textbf{0.7064} & 1.93\% \\
\texttt{\#17}&0.6739&0.6728 & 0.6775 & 0.6947 & 0.6953 & 0.6981 & \underline{0.6998} & \textbf{0.7027} & 1.45\% \\
\texttt{\#18}&0.6242&0.6280 & 0.6300 & 0.6394 & 0.6395 & 0.6412 & \underline{0.6417} & \textbf{0.6443} & 1.83\% \\
\texttt{\#19}&0.6193&0.6236 & 0.6219 &0.6465 & 0.6462 & 0.6498 & \underline{0.6505} & \textbf{0.6542}  & 2.46\%\\
\texttt{\#20}&0.6287&0.6271 & 0.6346 & 0.6437 & 0.6434 & 0.6487 & \underline{0.6503} & \textbf{0.6547} & 2.93\% \\
\midrule
\texttt{Overall}&0.6729&0.6763 & 0.6788 & 0.6864 & 0.6883 & 0.6922 & \underline{0.6954} & \textbf{0.7001} & 2.41\% \\
    \bottomrule
  \end{tabular}}
\end{table}

\subsection{Ablation Study (Q2)}
To investigate the effectiveness of each component in the proposed CSMN, we conduct several ablation experiments.

\subsubsection{The effectiveness of UIPN}
UIPN is devised to extract users' scenario-specific interests from their historical behaviors with respect to the target item and the visiting scenario simultaneously. Here, we devise two variant models including:
\begin{itemize}
    \item \textbf{CSMN w/o UIPN + T attention}: It removes UIPN while employing attention mechanism to extract users' interests from their behaviors only with respect to the Target item (T), ignoring the scenario information.
    
    \item \textbf{CSMN w/o UIPN + TS attention}: It removes UIPN while leveraging two specific attention modules to re-weigh users' historical behaviors with respect to Target item and the visiting Scenario (TS), respectively, followed by summarizing users' behaviors by weighted-sum to obtain users' interests.
\end{itemize}


\begin{table}[]
\small
    \caption{The effectiveness of UIPN.}
    \setlength\tabcolsep{3.0pt}
    \centering
    \label{tab:uipnablation}
    \begin{tabular}{c|cc}
    \toprule
    Model & AUC & RI\\
    \midrule
    CSMN & \textbf{0.7001} & 0.00\\
    CSMN w/o UIPN + T attention & 0.6952 & -2.51\%\\
    CSMN w/o UIPN + TS attention  & 0.6979  & -1.11 \%\\
  \bottomrule
    \end{tabular}
\end{table}

As shown in Table \ref{tab:uipnablation}, CSMN achieves the best performance compared with the other two variants. For example, compared with CSMN, \textbf{CSMN w/o UIPN + T attention} observes a performance drop of 2.51\% RI, which demonstrates the importance of extraction of users' interests with respect to different scenarios. \textbf{CSMN w/o UIPN + TS attention} observes a performance drop of 1.11\% RI, which demonstrates the effectiveness of eliminating the scenario-irrelevant information in users' behaviors with respect to the visiting scenario to further refine the representations of users' scenario-specific interests. 

\subsubsection{The effectiveness of URMN} To demonstrate the effectiveness of URMN, we remove it from the model, resulting in a variant model named \textbf{CSMN w/o URMN}. As shown in Table \ref{tab:urmnablation}, \textbf{CSMN w/o URMN} observes a performance drop of 1.63\% RI compared with CSMN, which demonstrates CSMN can effectively alleviate the cold-start issue attributing to URMN. Furthermore, we argue that the more sparse users' behaviors are, the greater relative improvement CSMN achieves. As shown in Table \ref{tab:overallresult}, CSMN gets the largest RI improvement over other competitors in scenario \#10 and scenario \#20, which have the largest portion of cold-start users.


\begin{table}[t]
\small
    \caption{The effectiveness of URMN.}
    \setlength\tabcolsep{3.0pt}
    \centering
    \label{tab:urmnablation}
    \begin{tabular}{c|cc}
    \toprule
    Model & AUC & RI\\
    \midrule
    CSMN & \textbf{0.7001} & 0.00\\
    CSMN w/o URMN & 0.6969 & -1.63\%\\
  \bottomrule
    \end{tabular}
\end{table}


\subsection{Parameter Sensitivity (Q3)}

To further understand the adverse effect of the cold-start issue in OTPs, we investigate the influence of the key hyper-parameters related to the issue, $i.e.$, the number of memory slots $q$, $\alpha _{k}$ (resp. $\alpha _{v}$) controlling the update rate of each memory slot's key (resp. value). First, as shown in Table~\ref{pic:sens}, CSMN achieves the best performance when $q$ takes the value of 1,000. Intuitively, the smaller $q$ is, the fewer the numbers of formed users' clusters are, and vice versa. Taking two extreme cases as example, one is $q=1$, where all the users have the same augmented interest vector, resulting in the difficulty of distinguishing different users' interests, especially for those with sparse behaviors. The other is an extremely large $q$, $i.e.$, $q=10,000$, where each slot of the memory will be too spare to update sufficiently. Next, we conduct seven groups of experiments, where each experiment setting the value of $\alpha _{k}$ and $\alpha _{v}$ as the same. As depicted in Table~\ref{pic:uprate}, CSMN obtains the best performance when $\alpha _{k}$ and $\alpha _{v}$ take the value of 0.3. Intuitively, when taking 0.0 as the update rate, $Mem_{j}^{Key}$ and $Mem_{j}^{Value}$ always follow the initialized values, neglecting the fact that parameters are continuously updating during model training, while taking as the update rate 1.0, $Mem_{j}^{Key}$ and $Mem_{j}^{Value}$ always taking the latest representation, abandoning the accumulated representation before. Obviously, both situations could not achieve the best performance, confirming that suitable updating rate is promising.

\begin{table}[t]
\small
    \caption{The effectiveness of memory size $q$.}
    \setlength\tabcolsep{3.0pt}
    \centering
    \label{pic:sens}
    \begin{tabular}{c|cccc}
\toprule
Memory Size & 10 & 100 & 1,000 & 10,000  \\
        \midrule
AUC & 0.6962 & 0.6979  & \textbf{0.7001}  & 0.6956  \\
\bottomrule
    \end{tabular}
\end{table}

\begin{table}[t]
\small
    \caption{The effectiveness of update rate $\alpha _{k}$ and $\alpha _{v}$.}
    \setlength\tabcolsep{3.0pt}
    \centering
    \label{pic:uprate}
    \begin{tabular}{c|ccccccc}
\toprule
Update Rate & 0.0 & 0.1 & 0.2 & 0.3 & 0.5 & 0.8 & 1.0  \\
        \midrule
AUC & 0.6947 & 0.6964  & 0.6982 & \textbf{0.7001} & 0.6971& 0.6955 & 0.6950 \\
\bottomrule
    \end{tabular}
\end{table}



\subsection{Online A/B Test (Q4)}

To further demonstrate the effectiveness of the proposed CSMN, we deploy it on our travel platform for A/B test, where the \textbf{Base} model is SAR-Net ~\cite{shen2021sar} and the evaluation metric is online CTR, $i.e.$, the number of clicks over the number of impression items. To make the online evaluation fair, confident, and comparable, both methods includes same number of users, $e.g.$, millions of users. We find the proposed CSMN achieves consistent improvement over SAR-Net model in consecutive seven days, $e.g.$, achieving an average improvement of 3.85\% CTR. To go a step forward, we find a more significant improvement is observed as was expected, $e.g.$, achieving an average improvement of 4.62\% CTR, for cold-start users, further demonstrating the effectiveness of CSMN dealing with cold-start issue. In a nutshell, the online A/B test results again demonstrate the effectiveness and practicability of our CSMN model in the industrial setting. Now, CSMN has been deployed on our platform and is serving all the traffic of twenty travel scenarios simultaneously.

\section{Conclusions}

In this paper, we propose a novel method named CSMN to deliver the unified click-through rate prediction task among the multiple scenarios on the online travel platforms. Specifically, it consists of two basic components including a User Interest Projection Network (UIPN) and a User Representation Memory Network (URMN), which can mitigate the cold-start issue effectively by exploiting scenario-shared and scenario-specific information of various scenarios simultaneously. Extensive experiments on both real-world offline dataset and online A/B test demonstrate the superiority of CSMN over state-of-the-art methods. How to employ the principle of meta-learning framework to further exploit the \emph{commonality} and \emph{discrimination} properties in multi-scenario CTR prediction task is an interesting topic and deserves more research efforts.

\subsubsection{Acknowledgments.}This work is supported by National Key Research and Development Program of China under Grant 2020AAA0107400 and National Natural Science Foundation of China (Grant No: 62206248).

\bibliographystyle{splncs04}
\bibliography{mybibliography}
%




\end{document}